\documentclass{amsart}

\begin{document}
\def\beq{\begin{eqnarray}}
\def\eeq{\end{eqnarray}}
\newcommand{\nn}{\nonumber}
\newcommand{\reals}{\mbox{${\rm I\!R }$}}
\newcommand{\nats}{\mbox{${\rm I\!N }$}}
\newcommand{\intgs}{\mbox{${\rm Z\!\!Z }$}}
\newcommand{\ujk}{u_{j,k}(1)}
\newcommand{\ujik}{u_{j,ik}(1)}
\newcommand{\uek}{u_{1,k}(1)}
\newcommand{\uzk}{u_{2,k}(1)}
\newcommand{\ueik}{u_{1,ik}(1)}
\newcommand{\uzik}{u_{2,ik}(1)}
\newcommand{\tk}{\frac{d}{dk}}
\newcommand{\pk}{\frac{\partial}{\partial k}}
\newcommand{\uja}{u_{j,k} (0)}
\newcommand{\ujb}{u_{j,k} (1)}
\newcommand{\vja}{v_{j,k} (0)}
\newcommand{\vjb}{v_{j,k} (1)}
\newcommand{\ujap}{u_{j,k}' (0)}
\newcommand{\ujbp}{u_{j,k}' (1)}
\newcommand{\ujx}{u_{j,k} (x)}
\newcommand{\vjx}{v_{j,k} (x)}
\newcommand{\ujxp}{u_{j,k}' (x)}
\newcommand{\vjxp}{v_{j,k}' (x)}
\newcommand{\ujxe}{u_{j,k}^{(1)} (x)}
\newcommand{\ujxz}{u_{j,k}^{(2)} (x)}
\newcommand{\ujxep}{u_{j,k}^{(1)\prime} (x)}
\newcommand{\ujxzp}{u_{j,k}^{(2)\prime} (x)}
\newcommand{\vjxe}{v_{j,k}^{(1)} (x)}
\newcommand{\vjxz}{v_{j,k}^{(2)} (x)}
\newcommand{\vjxep}{v_{j,k}^{(1)\prime} (x)}
\newcommand{\vjxzp}{v_{j,k}^{(2)\prime} (x)}
\newcommand{\drel}{\det (M+N H_{1,k} (1))}
\newcommand{\sca}{k^2 \langle u_{1,0} | u_{1,k} \rangle }
\newcommand{\scaneu}{k^2 \langle u_{1,0} | v_{1,k} \rangle }

\title{Functional determinants in the presence of zero modes}

\author{Klaus Kirsten}

\address{KK: Mathematics Department, Baylor University,
Waco TX 76798, USA\\
E-mail: Klaus\_Kirsten@Baylor.edu}

\author{Alan J. McKane}

\address{AJMcK: Department of Theoretical Physics,
University of Manchester \\Manchester M13 9PL,
 England\\
 E-mail: alan.mckane@man.ac.uk}


\begin{abstract}{We present a simple and accessible method which uses
contour integration methods to derive formulae for functional
determinants. To make the presentation as clear as possible we
illustrate the general ideas using the Laplacian with Dirichlet
boundary conditions on the interval. Afterwards, we indicate how
more general operators as well as general boundary conditions can
be covered.}
\end{abstract}

\maketitle

\section{Introduction}
In the Euclidean path-integral formalism, physical properties of
systems are conveniently described by means of path-integral
functionals. The leading order contribution to path-integrals
often involves the integration of the exponential of a quadratic
form. This integration can be carried out exactly and yields a
functional determinant\cite{fey65,sch81}.

In recent years, a contour integration method for the evaluation
of determinants has been developed\cite{bor96a,bor96b,KKbook}.
Although the applicability of the method in higher dimension seems
to be restricted to highly symmetric configurations, it turns out
that in one dimension quite generic situations can be
considered\cite{amkk}. In particular, the case with zero modes
present can be dealt with very elegantly and the determinants with
the zero mode extracted can be found in closed form.

We will first introduce the contour integral methods using the
simple example of the Laplace operator with Dirichlet boundary
conditions on the interval. There will be no zero modes present
and the example serves to introduce and explain some details of
the method. Based on the example, more general second order
operators with Dirichlet boundary conditions will be considered.
The results found can be compared with known
answers\cite{alt12,alt16}. Afterwards, we explain the additional
complications when zero modes are present. The conclusions
summarize the main results and will indicate further possible
generalizations of the approach to Sturm-Liouville type operators
with arbitrary boundary conditions.

\section{Example for the Laplacian on the interval}
\label{examples} Let us consider first the Laplacian
$L=-\frac{d^2}{dx^2}$ on the interval $I=[0,1]$ with Dirichlet
boundary conditions. The eigenvalues $\lambda_n^2$ and
eigenfunctions $u_n$ are defined by the equations \beq -\frac
{d^2}{dx^2} u_n (x) &=& \lambda_n^2 u_n (x) , \label{con1}\\
 u_n (0) &=& u_n (1) =0 . \label{con2}
 \eeq
 Equation (\ref{con1}) is satisfied by
 \beq
 u_n (x) = C_1 \cos (\lambda_n x) + C_2 \sin (\lambda_n x) .
 \label{con3}
 \eeq
 The boundary condition $u_n (0) =0$ forces $C_1 =0$, and $u_n (1)
 =0$ forces
 \beq
 \sin \lambda_n =0 \quad \Leftrightarrow \quad \lambda _n = \pi n
 , \quad n \in\intgs.\nn\eeq
The set of nontrivial linearly independent eigenfunctions is given
by $n\in\nats$. For this case where eigenvalues are known
explicitly, it is straightforward to write down the zeta function
associated with $L$, \beq \zeta_L (s) = \sum_{n=1} ^\infty \frac 1
{(\pi n)^{2s}} = \pi^{-2s}\zeta _R (2s) ,\nn\eeq and to calculate
the functional determinant in the zeta function
scheme\cite{ray71-7-145,hawk77-55-133,dowk76-13-3224}, \beq
-\ln\det L = \zeta _L ' (0) = -2 \zeta _R (0) \ln \pi +2 \zeta _R
' (0) = \ln \pi - \ln (2\pi )= - \ln 2 .\label{con1a}\eeq However,
for cases where (\ref{con1}) contains a potential, eigenvalues are
mostly not known and a simple procedure such as the one above is
not available. To make this more general case accessible, we will
therefore introduce contour integration methods to study the above
example. The methods are then directly generalizable to cases with
a potential.

Starting point for the contour integration method is the implicit
eigenvalue equation \beq \sin \lambda_n =0. \label{con4}\eeq This
equation allows the zeta function to be written as the contour
integral \beq \zeta _L (s) = \frac 1 {2\pi i} \int\limits_\gamma
dk \,\, k^{-2s} \frac \partial {\partial k} \ln \sin k ,\nn\eeq
where the contour $\gamma$ is counterclockwise and encloses all
eigenvalues on the positive real axis. As given, the
representation is valid for $\Re s > 1/2$. Next, one would like to
shift the contour towards the imaginary axis. In doing so, we pick
up a contribution from the origin because $\sin 0 =0$. Therefore,
a slightly more elegant formulation is obtained when starting with
\beq \zeta _L (s) = \frac 1 {2\pi i} \int\limits_\gamma dk \,\,
k^{-2s} \frac \partial {\partial k} \ln \frac{\sin k} k
,\label{con5}\eeq where the contribution from the origin is
avoided because now $\lim_{k\to 0} \frac{\sin k} k =1$. So now,
nothing prevents us from shifting the contour to the imaginary
axis and by taking due care of the cut of the complex root, which
we place along the negative real axis, we find \beq \zeta _L (s) =
\frac{ \sin \pi s} \pi \int\limits_0^\infty dk k^{-2s} \frac
\partial {\partial k} \ln \frac{ \sinh k} k .\label{con6}\eeq As
the $k\to 0$ behavior, and the $k\to \infty$ behavior, obtained
from \beq \frac{\sinh k} k = \frac{ e^k} {2k} \left(
1-e^{-2k}\right),\nn\eeq show, this integral representation is
well defined for $1/2<\Re s <1$. The restriction $\Re s < 1$ comes
from the $k\to 0$ behavior and $\Re s > 1/2$ comes from the $k\to
\infty$ behavior. Writing for some $\epsilon > 0$ \beq \zeta _L
(s) &=& \frac{ \sin \pi s} \pi \int\limits_0^\epsilon dk k^{-2s}
\frac
\partial {\partial k} \ln \frac{ \sinh k} k+ \frac{ \sin \pi s}
\pi \int\limits_\epsilon^\infty dk k^{-2s} \frac \partial
{\partial k} \ln \frac{ \sinh k} k,\nn\eeq we need to construct
the analytical continuation of the second integral to $s=0$. This
is achieved by adding and subtracting the leading $k\to\infty$
asymptotics, thus writing \beq \zeta _L (s) &=& \frac{ \sin \pi s}
\pi \int\limits_0^\epsilon dk k^{-2s} \frac \partial {\partial k}
\ln \frac{
\sinh k} k\nn\\
&+& \frac{ \sin \pi s} \pi \int\limits_\epsilon^\infty dk k^{-2s}
\frac
\partial {\partial k} \ln \left( \frac{ \sinh k} k 2k e^{-k}\right)
+ \frac{ \sin \pi s} \pi \int\limits_\epsilon^\infty dk k^{-2s}
\frac
\partial {\partial k} \ln \left( \frac{e^{k}} {2k} \right)\nn\\
&=& \frac{ \sin \pi s} \pi \int\limits_0^\epsilon dk k^{-2s} \frac
\partial {\partial k} \ln \frac{ \sinh k} k\nn\\
 &+& \frac{ \sin
\pi s} \pi \int\limits_\epsilon^\infty dk k^{-2s} \frac
\partial {\partial k} \ln \left( \frac{ \sinh k} k 2k e^{-k}\right)
+\frac{ \sin \pi s} \pi \left( \frac {\epsilon^{1-2s}} {2s-1} -
\frac {\epsilon^{-2s}} {2s} \right) ,\nn
 \eeq
valid around $s=0$.

The derivative can now be easily carried out and its value at
$s=0$ is given in its simplest form by letting $\epsilon \to 0$.
We obtain \beq \zeta _L ' (0) &=& \left.\lim_{\epsilon \to
0}\left\{ \ln \left( \frac{ \sinh k} k \right)\right| _0 ^\epsilon
+ \left. \ln \left( \frac{ \sinh k} k
2ke^{-k}\right) \right|_\epsilon^\infty \,\, -\epsilon + \ln \epsilon \right\} \label{ex1} \\
 &=& -\lim_{k\to 0} \ln \left( \frac{\sinh k} k\right) - \ln 2
 = -\ln 2 ,\nn\eeq
as given in (\ref{con1a}).

\section{More general second order operators}
Generalizing from the previous example, it is the aim of this
section to develop the contour integration approach for operators
of the type \beq L = -\frac {d^2} {dx^2} + V(x) \label{con21},
\eeq again on the interval $I=[0,1]$. The potential $V(x)$ is
assumed to be continuous and to start with all eigenvalues will be
assumed to be positive, so no zero modes are present. With an eye
on equation (\ref{con5}) we let $u_k (x)$ be the unique solution
of \beq (L - k^2) u_k (x) =0\label{con22}\eeq satisfying \beq u_k
(0) = 0, \quad u' _k (0) =1.\label{con23}\eeq The condition $u_k
(0) =0$ is chosen such that the eigenvalues are fixed by imposing
\beq u_k (1) =0.\label{con24}\eeq Then, the contour integral
representation of the associated zeta function is \beq \zeta _L
(s) = \frac 1 {2\pi i} \int\limits_\gamma dk \,\, k^{-2s} \frac
\partial {\partial k} \ln  u _k (1) .\label{intrep}\eeq The normalization $u'
_k (0) =1$ is chosen such that when shifting the contour towards
the imaginary axis no contributions from the origin arise, see
(\ref{con5}). As the leading $k\to\infty$ asymptotics does not
depend on the potential $V(x)$, the subsequent steps are precisely
as for the example in Section 2. Therefore, a glance at equation
(\ref{ex1}) shows that \beq \zeta _L ' (0) &=& -\lim_{k\to 0} \ln
\left( u_{ik} (1) \right) - \ln 2 = -\ln u_0 (1) - \ln 2 = -\ln
\left( 2u_0 (1)  \right),\label{gen1}\eeq where, by definition,
$u_0 (x)$ is the solution of the homogeneous differential equation
\beq L u_0 (x) =0 \quad \mbox{with}\quad u_0 (0) =0, \quad u' _0
(0) =1.\nn\eeq

\section{Presence of zero modes}
For the case when zero modes are present, the formula (\ref{gen1})
can not be applied simply because, as a zero mode, $u_0 (x)$
satisfies $u_0 (1) =0$. Therefore, the contour cannot be shifted
to the imaginary axis; instead, particular attention is needed
when approaching the origin in the $k$-plane. We first need to
determine the behavior of $u_{k}(1)$ for small $k$ in order to
eliminate the pole in the integrand. To do this, we note that
integrating the left-hand side of \beq \int^{1}_{0} dx\,u_{0}(x) L
u_{k}(x) = k^{2} \int^{1}_{0} dx\,u_{0}(x) u_{k}(x)\equiv k^{2}
\langle u_{0} | u_{k} \rangle   \nn \eeq by parts gives \beq
\left[ u'_{0}(x) u_{k}(x) - u'_{k}(x) u_{0}(x) \right]^{1}_{0} +
\int^{1}_{0} dx\,u_{k}(x) (L u_{0}(x)) = k^{2} \langle u_{0} |
u_{k} \rangle . \nn \eeq
%Here we have introduced the Hilbert space
%product $\langle \ | \ \rangle$ on ${\cal L}^2 (I)$ by \beq
%\langle u | v \rangle = \int _0 ^1 dx \,\, u(x)  \,\, v(x) . \nn
%\eeq
Using the boundary conditions this yields \beq u_{k}(1) =
\frac{k^{2} \langle u_{0} | u_{k} \rangle}{u'_{0}(1)} \equiv -
k^{2} f_{k} . \label{4} \eeq Since $f_{k}$ is finite and non-zero
as $k \rightarrow 0$, we have the desired behavior of $u_{k}(1)$.
The minus sign has been included in the definition of $f_{k}$ for
later convenience. It is important to note that (\ref{4}) is true
for any $k$.
% --- no small $k$ assumption was made to derive it.

We can now modify the discussion of Section 3 to cover the case
when a zero mode is present, by using the following observations.
The function $f_{k}$, defined by (\ref{4}), vanishes at all values
of $k$ for which $k^2$ is a positive eigenvalue, as $u_{k} (1)$
does. Only in the case of the zero eigenvalue do we have the
situation where $f_{0} \neq 0$, but $u_{0} (1)=0$. Thereforee if
we use $f_{k}$ instead of $u_{k} (1)$ in (\ref{intrep}), and
choose the contour so as to surround the positive eigenvalues, the
contour can be shifted towards the origin. However, the notion
that $u_{k} (1)$ may be straightforwardly substituted by $f_{k}$
has to be amended slightly, since these functions have different
behaviors as $k \to \infty$. This difference in behavior is
accounted for if we replace $u_{k} (1)$ not by $f_{k}$, but by
$(1-k^{2})f_{k}$, since then $u_{ik} (1) \sim k^{2}f_{ik}$ as $k
\to \infty$, while $(1-k^{2})f_{k}$ remains non-zero at $k=0$.

These remarks lead us to consider the contour integral \beq \frac
1 {2\pi i} \int_\gamma dk k^{-2s}\tk \ln (1-k^{2})f_{k}
=\zeta_{L_1} (s) + \frac 1 {2\pi i} \int_\gamma dk k^{-2s} \tk \ln
(1-k^{2}) \label{contour} \eeq where the contour surrounds all the
eigenvalues on the positive $k$ axis. Depending on whether the
contour encloses the point $k=1$, this might differ from $\zeta _L
(s) $ by a constant. But given our focus is on $\zeta _L ' (0)$,
this is irrelevant.
%This integral equals
%\beq \zeta_{L_1} (s) + \frac 1 {2\pi i} \int_\gamma dk k^{-2s} \tk
%\ln (1-k^{2}) , \label{contour_res} \eeq where it is understood
%that the zero mode has been omitted from the definition of the
%zeta-function. The second term in (\ref{contour_res}) is equal to
%$1$ if the contour surrounds $k=1$, and zero if it does not. Our
%final result does not depend on this choice so, for definiteness,
%we choose the contour to surround this point.

In summary, we are now in a position to proceed precisely as
before. In (\ref{gen1}), instead of having $u_0 (1)$ make its
appearance, its now $f_0$ that comes into play. The contributions
resulting from the asymptotics at infinity are not influenced at
all by the existence of a zero mode and therefore remain the same
as above in (\ref{gen1}). Therefore, again referring to this
equation, the answer now reads \beq \zeta _L ' (0) &=& -\lim_{k\to
0} \ln \left( f_k \right) - \ln 2= -\ln f_0 - \ln 2 = -\ln \left(
2f_0  \right).\label{gen1zero}\eeq For a particular problem
specified, namely for any given potential $V(x)$, the derivative
of the zero mode at the endpoint of the interval can be easily
determined numerically and so can the determinant.
\section{Conclusions}
The calculation provided can be generalized in various respects.
First, instead of considering (\ref{con21}), one can analyze
determinants for general Sturm-Liouville type operators \beq L= -
\frac d {dx} \left( P(x) \frac d {dx}\right) + V(x) .\nn\eeq
Reformulating the second order differential operator as a first
order system we then introduce $v_k (x)= P (x) u_k ' (x)$. General
boundary conditions are imposed as \beq M  {\uja \choose \vja } +
N { \ujb \choose \vjb} = {0 \choose 0} , \label{gbc} \eeq where
$M$ and $N$ are $2\times 2$ matrices whose entries characterize
the nature of the boundary conditions. For this quite generic
situation closed formulas for the determinant can be derived and
applied for different examples\cite{ajmkkprep}. We can also relax
the condition that the eigenvalues are nonnegative. In order that
the associated zeta function can be defined in the presence of
negative eigenvalues, we place the cut of the complex root at some
angle different from $0$ and from $\pi$. Choosing a suitable
contour $\gamma$ for the representation of the zeta function, only
minor changes to the procedure explained are
necessary\cite{ajmkkprep}.

\end{document}